\documentclass[a4paper]{aa}
\usepackage{graphicx}
\usepackage{txfonts}
\usepackage{amssymb}
\usepackage{natbib}
\bibpunct{(}{)}{;}{a}{}{,} 
\usepackage{bm}

\begin{document}
\title{Time-dependent CO depletion during the formation of protoplanetary disks}

\author{
C.~Brinch \and
R.~J.~van Weeren \and  
M.~R.~Hogerheijde}
\date{}
\institute{
  Leiden Observatory, Leiden University,
  P.O.~Box 9513, 2300 RA Leiden, The Netherlands\\
  \email{brinch@strw.leidenuniv.nl}
  }

\abstract
{Understanding the gas abundance distribution is essential when tracing star
 formation using molecular line observations. Changing density and temperature
 conditions cause gas to freeze-out onto dust grains, and this needs to be 
 taken into account when modeling a collapsing molecular cloud.}
{This study aims to provide a realistic estimate of the CO abundance 
 distribution throughout the collapse of a molecular cloud. We provide 
 abundance profiles and synthetic spectral lines which can be compared to 
 observations.}  
{We use a 2D hydrodynamical simulation of a collapsing cloud and subsequent 
 formation of a protoplanetary disk as input for the chemical calculations. 
 From the resulting abundances, synthetic spectra are calculated using a 
 molecular excitation and radiation transfer code.}
{We compare three different methods to calculate the abundance of CO. Our models
 also consider cosmic ray desorption and the effects of an increased CO binding 
 energy. The resulting abundance profiles are compared to observations from the 
 literature and are found to agree well.}
{The resulting abundance profiles agree well with analytic approximations, and
 the corresponding line fluxes match observational data. Our developed method to 
 calculate abundances in hydrodynamical simulations should greatly aid in 
 comparing these to observations, and can easily be generalized to include 
 gas-phase reaction networks.}

\maketitle

\section{Introduction}
The study of low-mass Young Stellar Objects (YSOs), from early protostellar 
cores to T Tauri stars surrounded by disks, involves observations of either the 
thermal emission from dust grains or molecular line emission. While dust 
emission yields information about the temperature profile
\citep[e.g.,][]{draine1984}, and, through measurements of the shape of the 
Spectral Energy Distribution, the evolutionary stage~\citep{lada1984,
adams1987}, spectral lines are the only way to constrain the kinematical 
properties of YSOs. The interpretation of molecular lines are crucial for the 
understanding of protoplanetary disk formation and the distribution of angular 
momentum during these stages of star formation~\citep{evans1999}. However, 
deriving the gas motion from spectral line profiles is complicated by 
degeneracies between the velocity field topology, inclination of the angular 
momentum axis, optical depth, and geometrical effects~\citep{brinch2007}. 

In addition to these effects, molecular abundances may vary greatly. The gas 
abundance is determined by the ongoing chemistry driven by the co-evolving 
density and temperature distributions, and to some extent also the velocity 
field. The variation in molecular abundances throughout a YSO may have a huge 
impact on line profiles and therefore it is important to understand the 
chemistry in order to interpret these line profiles. The effect of the 
chemistry is enhanced when observations at higher resolution are done. If the 
observations are done at a relatively low resolution ($\gtrsim$10$''$), the 
emission is smeared out over a large portion of the object and therefore an 
average constant abundance is mostly adequate. However, when going to a higher 
resolution, using sub-millimeter interferometry ($\sim$1$''$), the emission is 
averaged over a smaller part of the object and thus variations in the abundance 
need to be taken into account.

The problem of determining the abundance distribution has been addressed before
in the literature. \citet{lee2004} took a semi-analytical approach using a 
self-similar collapse of an isothermal sphere in which they followed 
``fluid-elements''. In that paper, the chemistry was followed in a series of 
Bonnor-Ebert spheres during the pre-stellar phase and in a self-similar inside-out
collapse during the protostellar phase. This model provides a good analytical 
description of the early stages of star formation, but, lacking any rotational 
velocity, no disk is formed in this scenario. \citet{aikawa2001,aikawa2003,
aikawa2005} followed the chemical evolution of contracting pre-stellar clouds 
using an approach similar to \citet{lee2004} but using an isothermal cloud
collapse only. 
Their calculations consider only the very early stages of the star formation 
process. \citet{jorgensen2002,jorgensen2005} introduced the ``drop model'', 
where chemical depletion occurs as a step function when certain temperature and 
density conditions of any underlying model are met. While this approach is 
simple yet quite successful, there may be cases where the abundance changes 
only very gradually so that a step function is no longer a good approximation.  
More recently, \citet{Tsamis2008} calculated HCO$^+$, CS, and N$_2$H$^+$ abundances
and spectra from a simulation of an inside-out collapsing core.

In this paper we investigate detailed evolution of the gas-phase and solid 
state abundance of CO in a collapsing, rotating core. We base our study on a 
hydrodynamical simulation, where we follow the CO freeze-out and evaporation 
for a number of test particles that flow with the gas. The lay-out of the paper 
is as follows: In Section~\ref{model} we present our numerical simulations and 
the equations used to solve the CO freeze-out chemistry. In Section 
\ref{results} we show three different chemical models and how our results 
compare to observations as well as the results of previous papers. Sections
\ref{discussion} and~\ref{summary} hold a discussion and a summary 
respectively.

\section{Tracing chemistry during disk formation}\label{model}
\subsection{The physical model}
Instead of using an analytic description of the gravitational collapse and
subsequent disk formation, we use a grid-based 2D axis-symmetric 
hydrodynamical scheme. The code which is used is described in 
\citet{yorke1999} and in particular we adopt the J-type model described in that 
paper.  
\begin{figure*}
  \begin{center}
  \includegraphics[width=8.5cm]{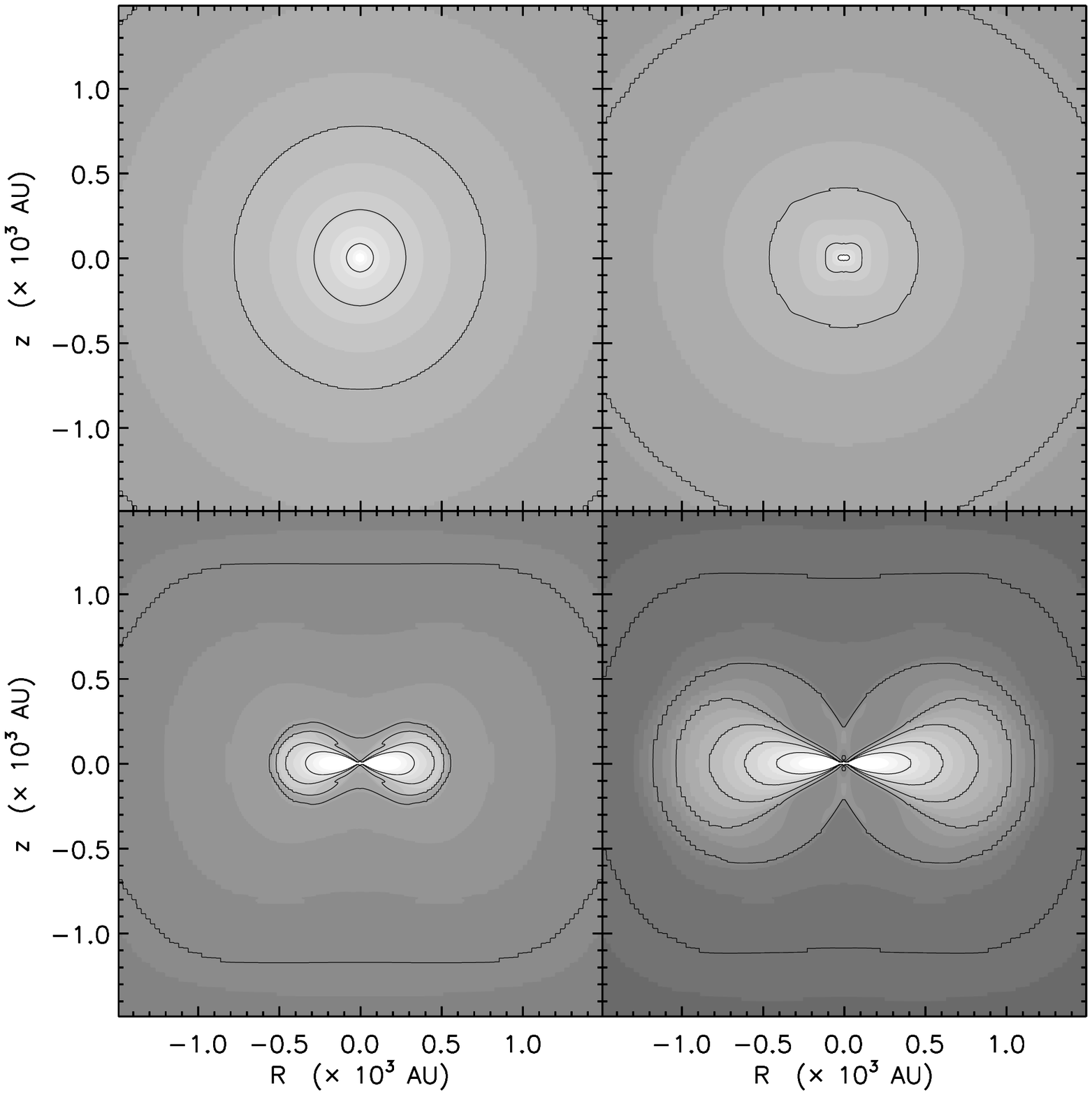}
  \includegraphics[width=8.5cm]{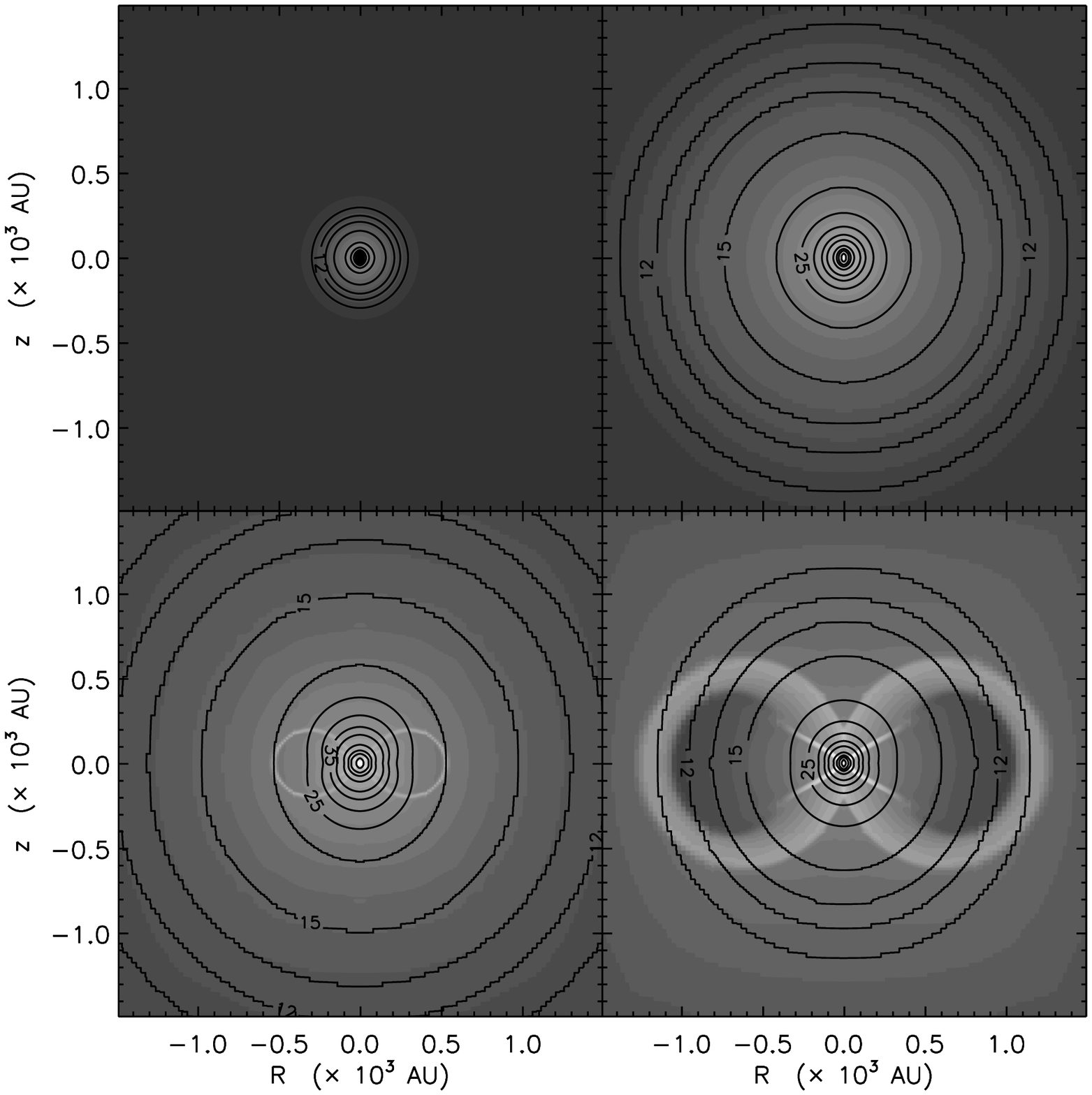}
  \caption{The density (left panel) and temperature (right 
 		   panel) contours of four time snapshots of the hydrodynamical 
		   simulation. Time progresses from left to right, top to bottom at 
		   t = \{0.0(6), 0.5, 1.5, 2.5\} t$_{ff}$. In the right panel, contour 
		   lines refer to the dust temperature, while the color scale denote 
		   the gas temperature.}\label{hydro-snaps}
\end{center}
\end{figure*}

The initial conditions for this model are a 1 M$_{\odot}$ isothermal sphere 
with a temperature of 10 K and a power-law density slope of $\rho \propto 
r^{-2}$. The cloud has an outer radius of 1$\times 10^{15}$ m (6667 AU), and an 
initial solid-body rotational perturbation of 1$\times 10^{-13}$ s$^{-1}$ is 
given. The model evolves under the action of gravity while the temperature is 
solved for self-consistently using an approximate radiation transfer method. 
Angular momentum is transferred through artificial viscosity using an 
$\alpha$-prescription~\citep{shakura1973}. 

The age of the simulated system is described in terms of the initial free-fall
timescale,
\begin{eqnarray}
  t_{ff}=\sqrt{\frac{\pi^2 R^3}{8GM},}
\end{eqnarray}
where $M$ is the mass of the cloud and $R$ is the radius. For the initial
conditions in our particular simulation one free-fall time is $1\times 10^5$ 
yr. We let the simulation run for about 2.6 $t_{ff}$ at which point the 
simulation terminates due to extreme velocities near the center. Figure
\ref{hydro-snaps} shows four time snapshots at characteristic ages. We use the 
same four snapshots throughout this paper, at times of 0.0, 0.5, 1.5, and 2.5 
t$_{ff}$ (with the exception of Fig.~\ref{hydro-snaps}, where we use a few 
snapshots later than t = 0.0 in order for the temperature to have evolved 
slightly from the isothermal initial condition).  

\begin{figure}
  \includegraphics[width=8.5cm]{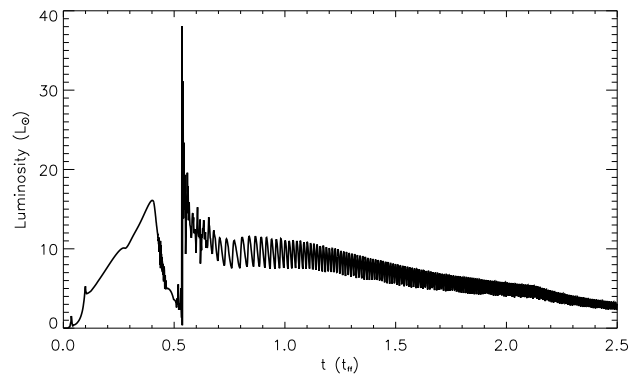}
  \caption{The time evolution of the luminosity of the central source in the
           hydrodynamical simulation. This figure compares to Fig.~7 in
		   \citet{yorke1999}.}\label{luminosity}
\end{figure}

The luminosity is given by the sum of the intrinsic stellar luminosity and
the accretion luminosity
\begin{eqnarray}
  L_{acc} = \frac{3}{4}\frac{GM_*\dot{M}}{R_*},
\end{eqnarray}
where $M_*$ is the mass of the star, $R_*$ the radius of the star, and 
$\dot{M}$ the mass flux onto the star. The total luminosity is shown, plotted as
a function of free-fall time, in Fig.~\ref{luminosity}. After the initial increase
in the luminosity, it drops for a while, after which it steeply increases again 
and then slowly falls of linearly for the remainder of the simulation. As 
described in~\citet{yorke1999}, the smooth increase and the drop around 0.5 
$t_{ff}$ is dominated by the accretion luminosity where low angular momentum 
material is able to fall directly onto the star. The sharp increase in luminosity 
marks the formation of an equilibrium disk and from then on, the intrinsic 
stellar luminosity dominates the total luminosity.

The hydrodynamical simulations does not include any chemistry and because of that, 
we cannot let the state of the molecules, whether they are in the gas phase or locked in an 
ice matrix, couple back to the hydrodynamics. We therefore determine the abundance 
by post-processing the result of the hydrodynamical calculations. We follow the 
chemistry by populating the computational domain of the first snapshot with 
trace particles. These particles are massless and do not interact with each 
other in any way. The trace particles are positioned evenly in the radial 
direction and $9\times 10^5$ particles are used. Every trace particle 
represents the local molecular environment and carries information on the state 
of the molecules (ice or gas). The particles are then allowed to follow the 
flow, predetermined by the hydrodynamical calculations, and the state of the 
particles is updated as temperature and density conditions change throughout 
the hydrodynamical simulation. Finally, the state of the trace particles are 
mapped back onto the hydro-grid at each output time step, so that a complete 
history from t=0.0 t$_{ff}$ to t=2.5 t$_{ff}$ of the CO abundance as function 
of $R$ and $z$ is linked to the density and temperature. Taking the CO 
distribution into account, we can use each of these time snapshots as input 
models for our 2D line excitation and radiative transfer code \emph{RATRAN} 
\citep{hogerheijde2000a} to predict line profiles of CO and its optically thin 
isotopologue C$^{18}$O. 

\begin{figure}
  \includegraphics[width=8.5cm]{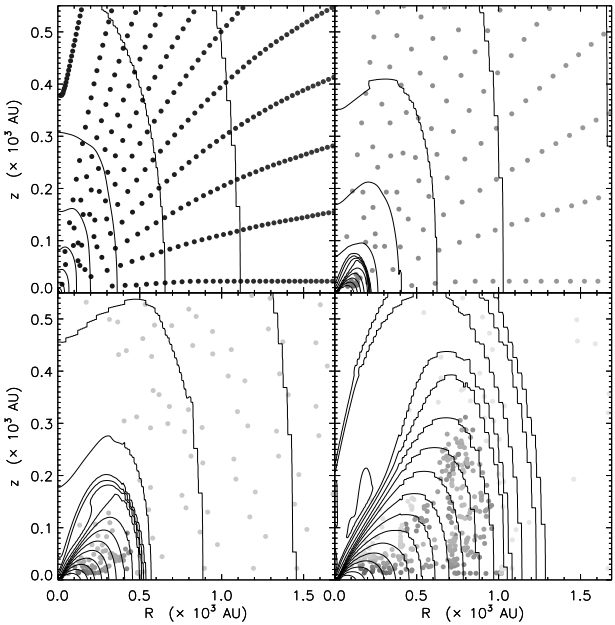}
  \caption{Trace particle motion in the hydrodynamical simulation. The 
  		   particles are color coded according to their initial radial 
		   position. Contour lines describe the density. The vertical color bar 
		   shows the particles initial distance from the center. The four panels (from the 
		   top-left to bottom-right) show the particle positions at t=0.0, 0.5, 
		   1.5, and 2.5 t$_{ff}$.}\label{particles}
\end{figure}

Before proceeding with the freeze-out calculations, it is interesting to 
consider the dynamics of the trace particles in order to develop an 
understanding of the environments that the particles traverse. By a simple 
radial color coding of the particles, we can effectively visualize the flow 
within the hydrodynamical simulation and track how material is accreted onto 
the disk. Four snapshots of the particle distribution are shown in Fig.
\ref{particles} where the shading of the particles refers to their initial 
distance from the center. The disk is seen to be layered vertically rather than 
radially which is somewhat counterintuitive. The vertical layering arises 
because material joins the disk from above. As the disk spreads viscously due 
to conservation of angular momentum, a shock front propagates outwards, pushing
aside infalling material. The only possible path of the particles is therefore 
upwards and over the disk lobe until they can rain down onto the disk at 
smaller radii. This rolling motion, is also reflected in the strong vertical 
mixing in the outer disk. However, this behavior may be an artifact related to 
the 2D nature of the hydrodynamical code. In a 3D simulation, particles would 
also be able to dissipate in the azimuthal direction, which would likely change 
their motions.

\subsection{Freeze-out and evaporation}
In general, molecular gas abundances depend on reaction rates and phase
transitions. However, under the physical conditions present in our model, only 
the latter has any significant impact on the CO abundance. The state of CO is 
governed by rate equations. The two main mechanisms are depletion, where gas 
phase molecules freeze-out onto grains, and desorption, where solid state 
molecules evaporate into the gas phase.

The depletion rate $\lambda$ used in this paper is given by the equation
\citep[e.g.,][]{Charnley2001},
\begin{eqnarray}\label{dep}
  \lambda = \pi a^2 \sqrt{\frac{8kT_g}{\pi m_{CO}}} n_{grain},
\end{eqnarray}
where $T_g$ is the gas temperature, $m_{CO}$ is the mass of a CO molecule, and 
$n_{grain}$ is the grain number density, assuming a grain abundance of 
1.33$\times$ 10$^{-12}$ relative to the hydrogen nucleon density. For the mean
grain size $a$ we adopt the value of 0.1$\mu$m. We also assume a sticking 
probability of unity. 

A similar equation can be written for the desorption of molecules. The thermal 
desorption rate is given by~\citet{watson1972},
\begin{eqnarray}\label{des}
  \xi_{th} = \nu(X) e^{\left(\frac{-E_b(X)}{k_b T_d}\right )},
\end{eqnarray}
where $k_b$ is Boltzmann's constant, $T_d$ is the dust temperature, $\nu(X)$
is the vibrational frequency of $X$ in its binding site~\citep{Hasegawa1992}, 
and $E_b(X)$ is the binding energy of species $X$ onto the dust grain. In this 
paper we use a default binding energy of 960 K, but other values, corresponding 
to mantles which are not composed of pure CO ice, are explored as well in Sect.
\ref{cr_ibe}.

A second desorption mechanism which is taken into account is cosmic ray induced
desorption. This mechanism was first proposed by~\citet{watson1972}. Energetic 
nuclei might eject molecules from grain surfaces by either raising the 
temperature of the entire grain or by spot heating near the impact site. The 
formulation of~\citet{Hasegawa1993} is used to calculate the cosmic ray
desorption rate,
\begin{eqnarray}\label{des:cr}
  \xi_{cr} = 3.16\times 10^{-19} \xi_{th} \Big \arrowvert_{T_d = 70 K}.
\end{eqnarray}
Cosmic ray desorption can be added to the thermal desorption, effectively
preventing a depletion of 100\% under any condition.

With the depletion and desorption rates, defining $f_d$ to be the normalized 
fractional depletion, we can write an equation for $f_d$,
\begin{eqnarray}\label{eq:abun}
\frac{df_d(\mathbf{r},t)}{dt} = \lambda(\mathbf{r},t)
f_d(\mathbf{r},t)-\xi(\mathbf{r},t)\big (1-f_d(\mathbf{r},t)\big ),
\end{eqnarray}
in which we implicitly use the boundary condition that the sum of CO molecules 
in solid state and in the gas phase is constant.

Equation~\ref{eq:abun} is a very general one and the rate functions can easily
be changed so that the equation describes other reactions. \citet{Weeren2007} 
shows that in principle a whole set of such equations can be solved 
simultaneously using gas phase reaction rates from e.g., the UMIST database
\citep{woodall2007} although computational demand becomes an issue for a full 
network. 

\section{Results}\label{results}
To obtain abundance distributions we evaluate Eq.~\ref{eq:abun} using the gas 
temperature $T_g$, the dust temperature $T_d$, and the mass density (we assume
that H$_2$ is the main mass reservoir and that the gas-to-dust ratio is 100:1) 
from the hydrodynamical simulation. We assume that the cloud has been in 
hydrostatic equilibrium prior to collapse for $3\times 10^5$ yr 
(=$1\times10^{13}$ s). The density profile before collapse is given by a 
spherical power-law with a uniform temperature as described in Sect.~\ref{model}. Starting with a pure gas phase abundance of $2\times 
10^{-4}$ molecules relative to H$_2$, we let the abundance evolve for $3\times 
10^5$ yr with a constant temperature of 10 K and a static density profile. The 
new CO abundance serves as initial condition for the collapse, i.e., the start 
of the hydrodynamical simulation which defines t = 0. In the following we 
present three different ways to obtain the abundance profiles.

\subsection{Model 1: Drop abundance}
This approach is denoted the drop abundance model, because it follows directly
from the method presented by~\citet{jorgensen2005}. In this approach we do not
actually solve Eq.~\ref{eq:abun}, but rather consider the rate
equations~\ref{dep} and~\ref{des}. The rate equations are inverted to obtain a
characteristic time scale, and at any given time in the simulation, this time
scale is evaluated based on the current density and temperature conditions. If
the freeze-out time scale is shorter than the evaporation time scale, the
gas phase abundance drops to a preset level (sufficiently low, i.e., several 
orders of magnitude) but if the actual age of the cloud is shorter than the 
freeze-out time scale, the gas phase abundance is kept at its initial value. 
Using this approach, the resulting profile is a step function where all CO 
is either in the gas phase or in the solid state. The movement of the trace 
particles does not affect the result since the abundance is determined only 
from the current local conditions. The effect of cosmic ray desorption is 
implicitly accounted for by not letting the abundance drop to zero, but only 
two or three orders of magnitude, although freeze-out conditions are met.
\begin{figure}
  \includegraphics[width=8.5cm]{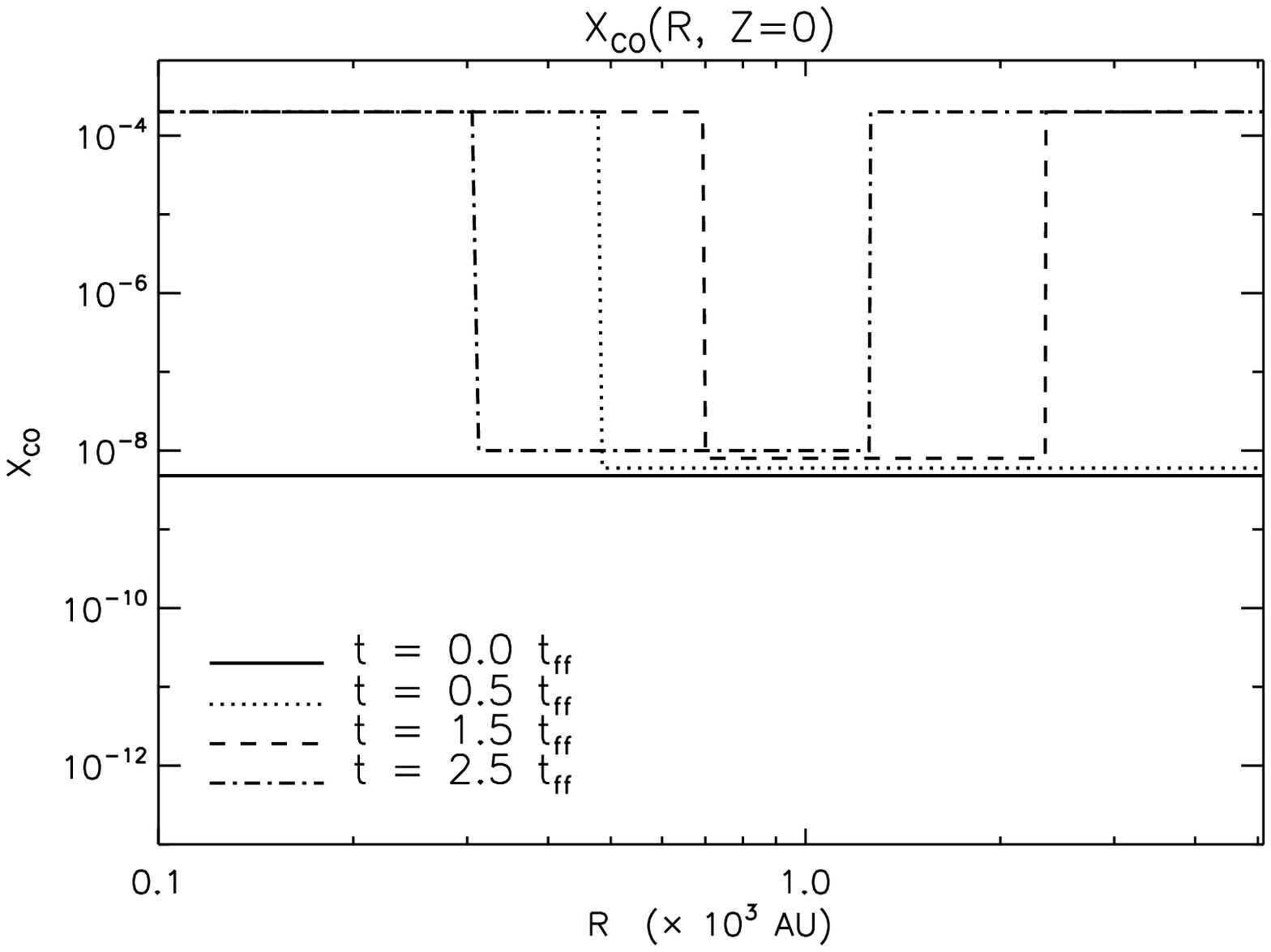}\\
  \includegraphics[width=8.5cm]{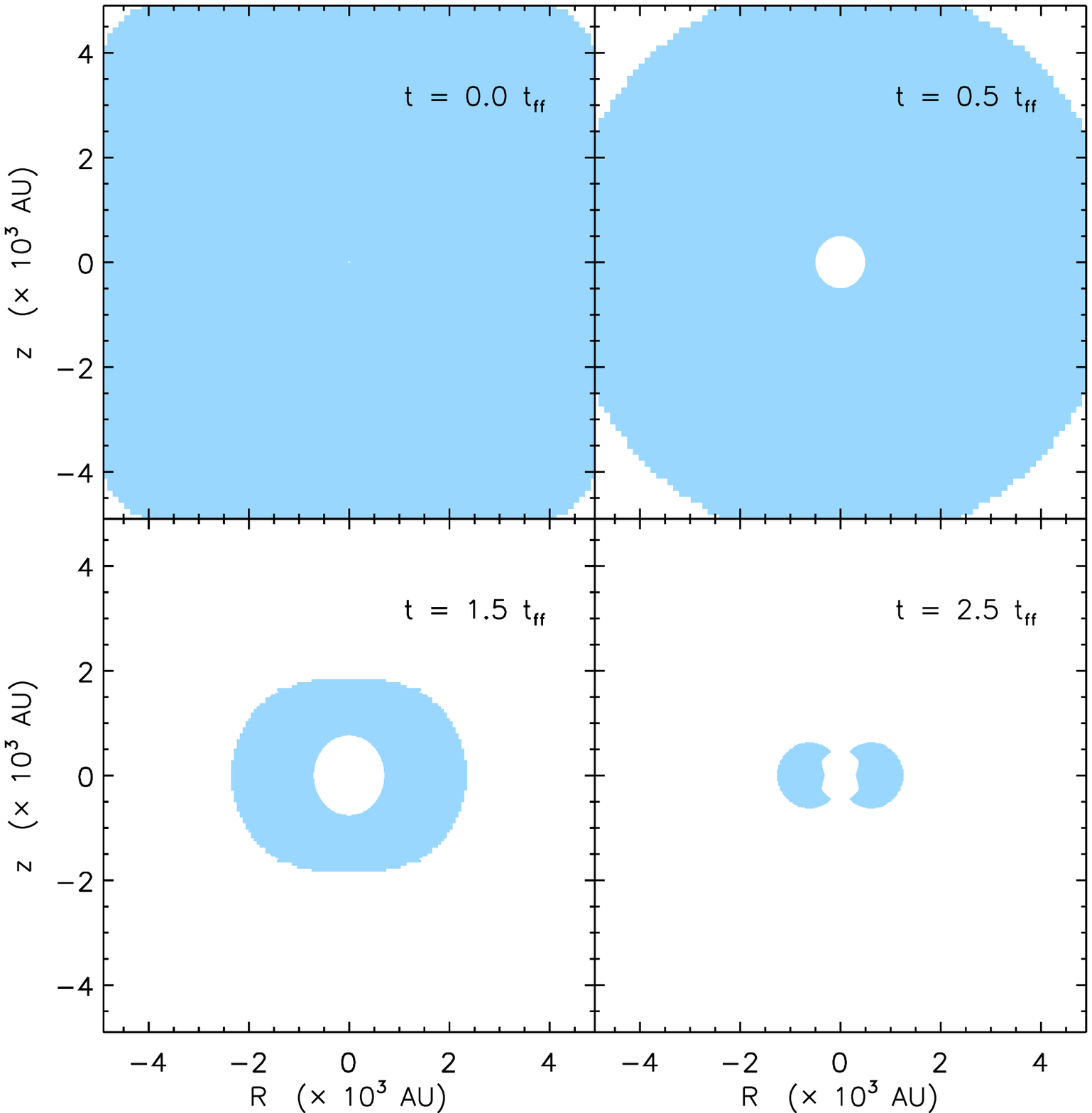}
  \caption{This figure shows the abundance distribution of Model 1. The top panel
           shows the radial abundance profile through the mid-plane of the disk.
		   The bottom panel shows the spatial abundance distribution. The shaded
		   area is where molecules are frozen out. The unit on the y-axis of the
		   top panel is fractional abundance.}\label{model1}
\end{figure}

The result is shown in Fig.~\ref{model1}. The top panel shows the radial 
abundance profile along the disk mid-plane ($z$=0), while the lower panel shows
the two dimensional distribution. The result of this drop abundance approach 
supports the idea of an evolution of the CO abundances from a pre-stellar core 
to a Class I object, proposed by~\citet{jorgensen2005}. The maximum size of 
our depleted zone is $\sim$6000 AU, just before the collapse sets in, and 
shrinks to $\sim$500 AU (when averaged in radial and polar directions) as the 
cloud collapses and the disk is formed. The inner evaporated zone has a radius 
of $\sim$1000 AU, in agreement with values derived in~\citet{jorgensen2005}. 

\subsection{Model 2: Variable freeze-out}
\begin{figure}
  \includegraphics[width=8.5cm]{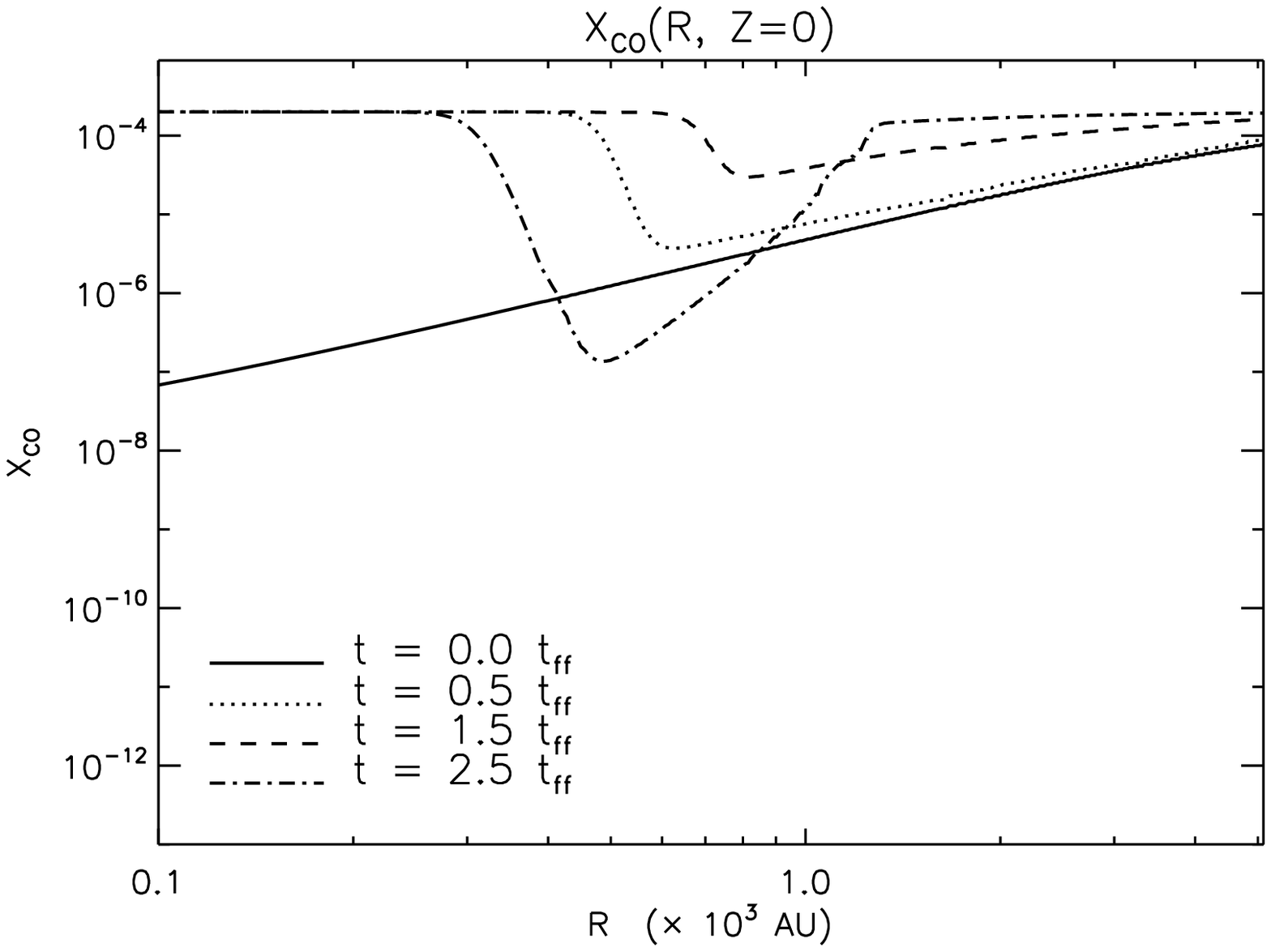}\\
  \includegraphics[width=8.5cm]{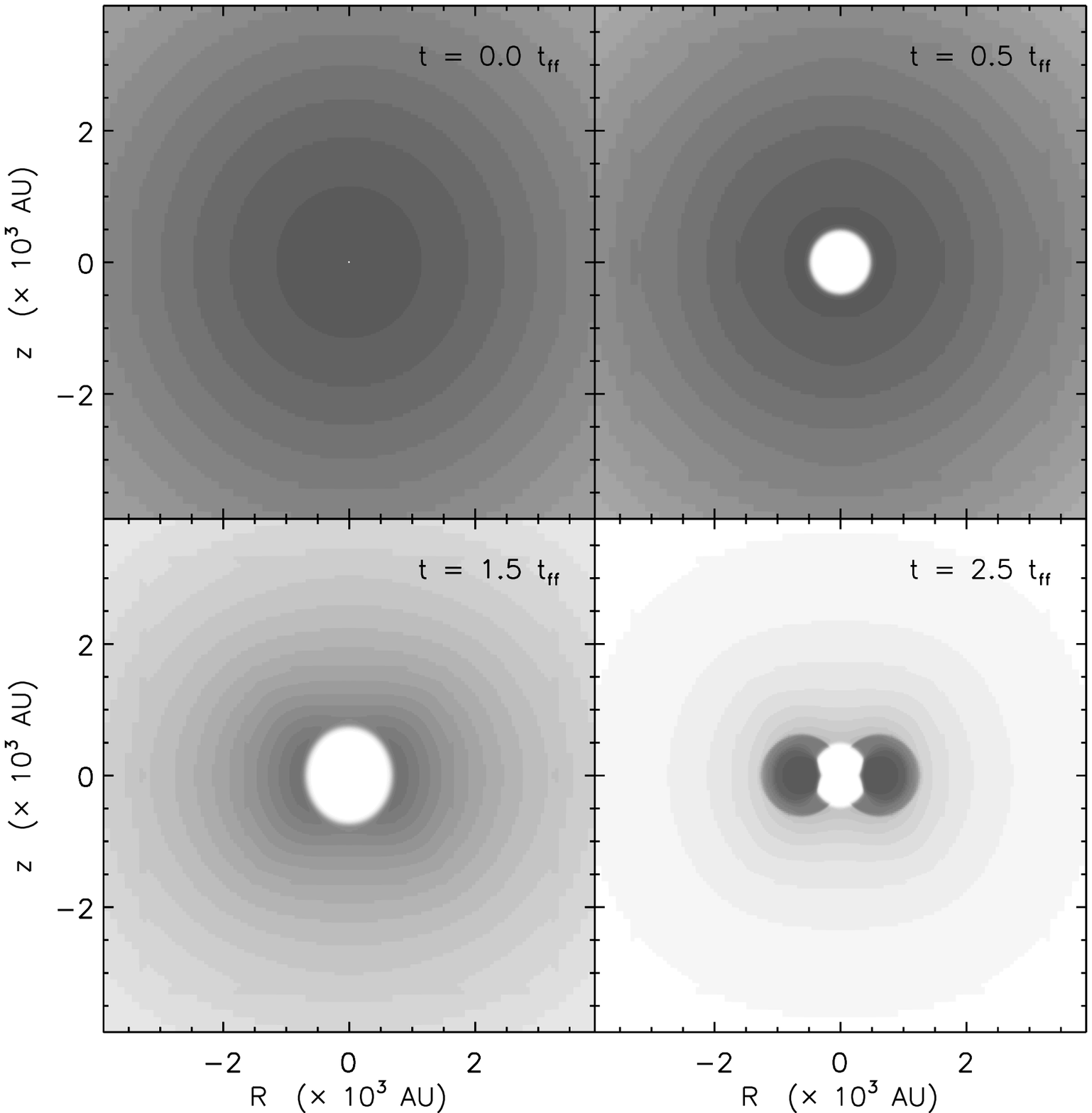}
  \caption{Abundance distribution of Model 2. Otherwise identical to 
   		   \ref{model1}.}\label{model2}
\end{figure}

The second method is similar to Model 1 except that instead of using the time
scales, we use the actual rate equations which allows the trace particles to be
in a mixed state. The values for $\lambda$ and $\xi$ are calculated for all 
$\mathbf{r}$ using the current $n(\mathbf{r})$ and $T(\mathbf{r})$ and 
Eq.~\ref{eq:abun} is solved to obtain the fraction of gas phase CO to solid 
state CO for each trace particle position. In this model, we still do not 
consider the history of the trace particles and thus the motion of the 
particles is irrelevant and the time dependence of $\xi$ and $\lambda$ in
Eq.~\ref{eq:abun} drops out. Plots of the abundance distributions are shown in 
Fig.~\ref{model2}. The profiles of Model 2 are quite similar to the ones from 
Model 1, except for the gradual change from pure gas phase to pure solid state. 
Especially the inner edge is well approximated by the step function of Model 1 
because of the exponential factor in the evaporation, described by Eq.
\ref{des}. The main difference between the two models is the outer edge, and 
especially at earlier times, where the gradient is more shallow.

\subsection{Model 3: Dynamical evolution} 
Our final model takes the motion of the trace particles and their chemical 
history into account. Freeze-out and evaporation rates are calculated 
dynamically for each time step and composition of the particles is used as 
initial condition for Eq.~\ref{eq:abun}. For example, the depletion in a 
certain area represented by a trace particle depends on where this particle has 
come from, which we know, because the trace particles follow the flow 
calculated in the hydrodynamical simulation. Two trace particles may pass by 
the same region at the same time and experience the same desorption and 
depletion conditions. However, they may have come from different areas, e.g., 
one may have spent time in a warmer region and the other may come from a colder 
region and therefore they will leave the place with different compositions. 

\begin{figure}
  \includegraphics[width=8.5cm]{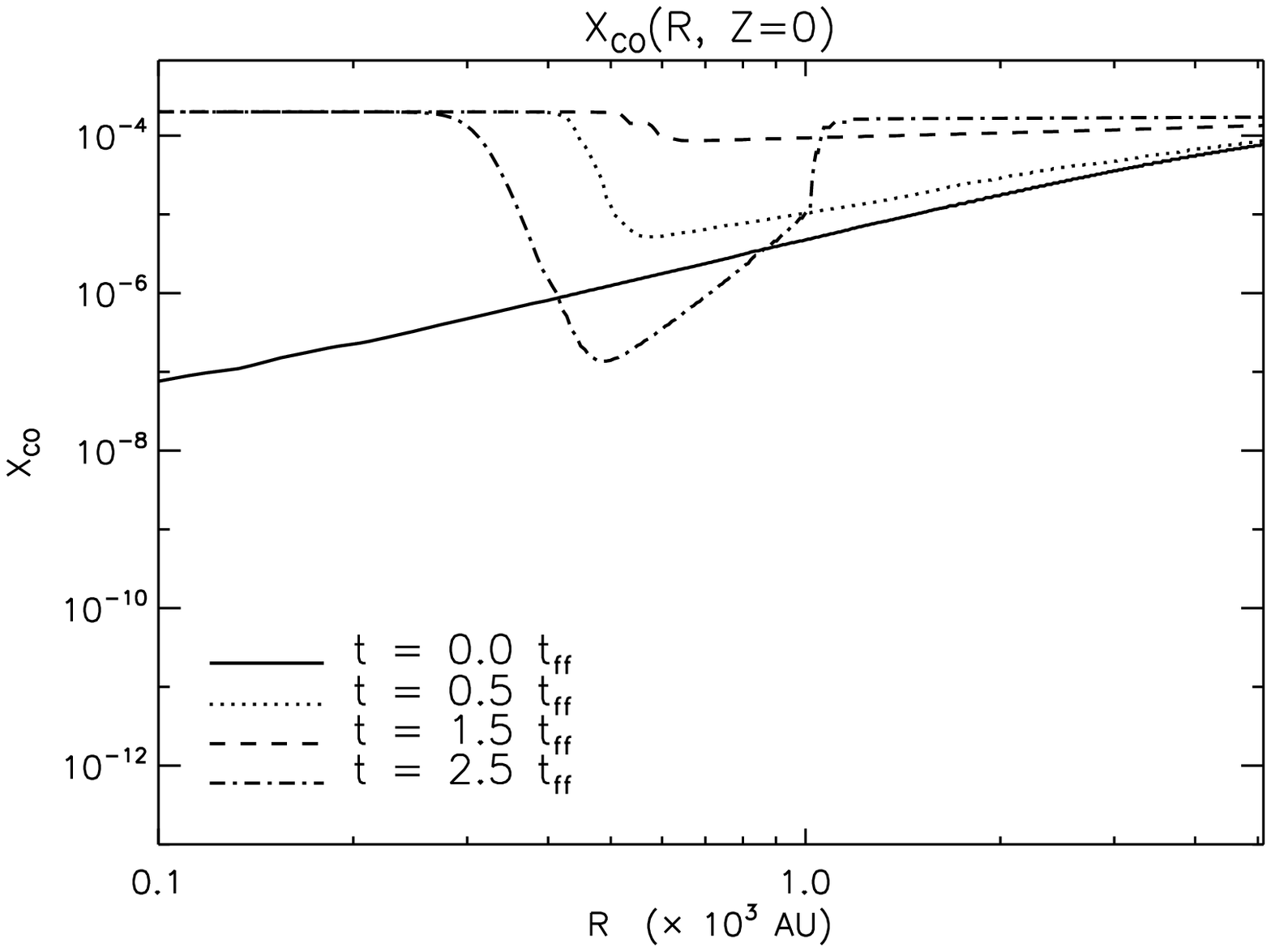}\\
  \includegraphics[width=8.5cm]{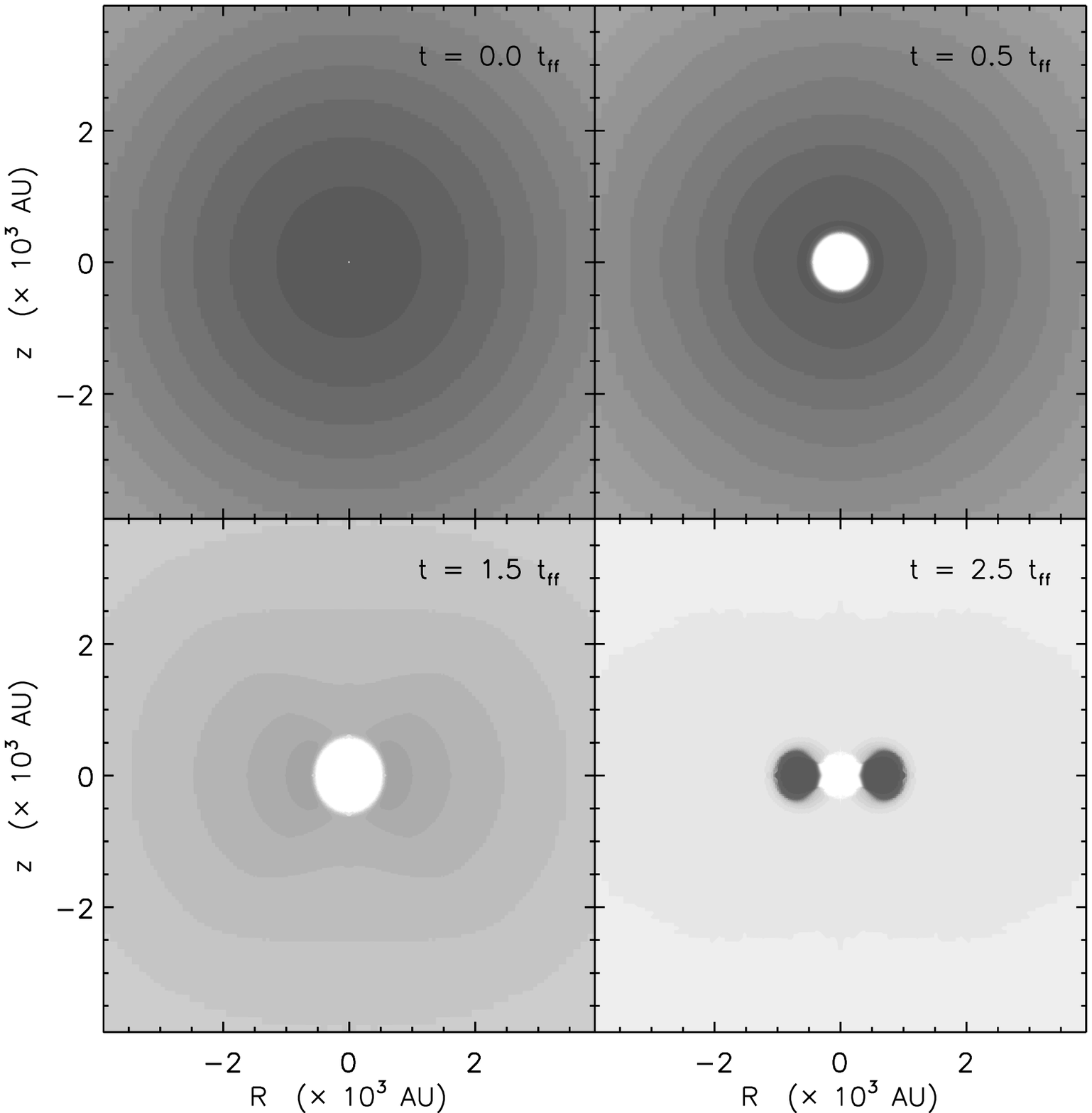}
  \caption{Abundance distribution of Model 3. Otherwise similar to~\ref{model1}.}\label{model3}
\end{figure}

Figure~\ref{model3} shows the abundance distributions for this model. 
Surprisingly, the profiles are not very different from those of Model 2. A 
difference only occurs when the dynamical time scale is comparable to the 
chemical time scale. In this case, a gas phase trace particle may move out
of a depletion zone faster than depletion can make any significant change to 
its composition, or vice versa. The evaporation front around the disk is also 
sharper compared to Model 2 which makes Model 3 appear almost like a step 
function in the latter part of the simulation.

\subsection{Cosmic ray desorption and increased binding energy}\label{cr_ibe}
We now investigate the effects of excluding the cosmic ray desorption term
(Eq.~\ref{des:cr}) to see what effect this mechanism has on the abundances. We
also increase the CO binding energy $E_b$ in Eq.~\ref{des} in order to emulate 
different ice mantle compositions. In the following we use Model 3 as a 
template.

\begin{figure}
  \includegraphics[width=8.5cm]{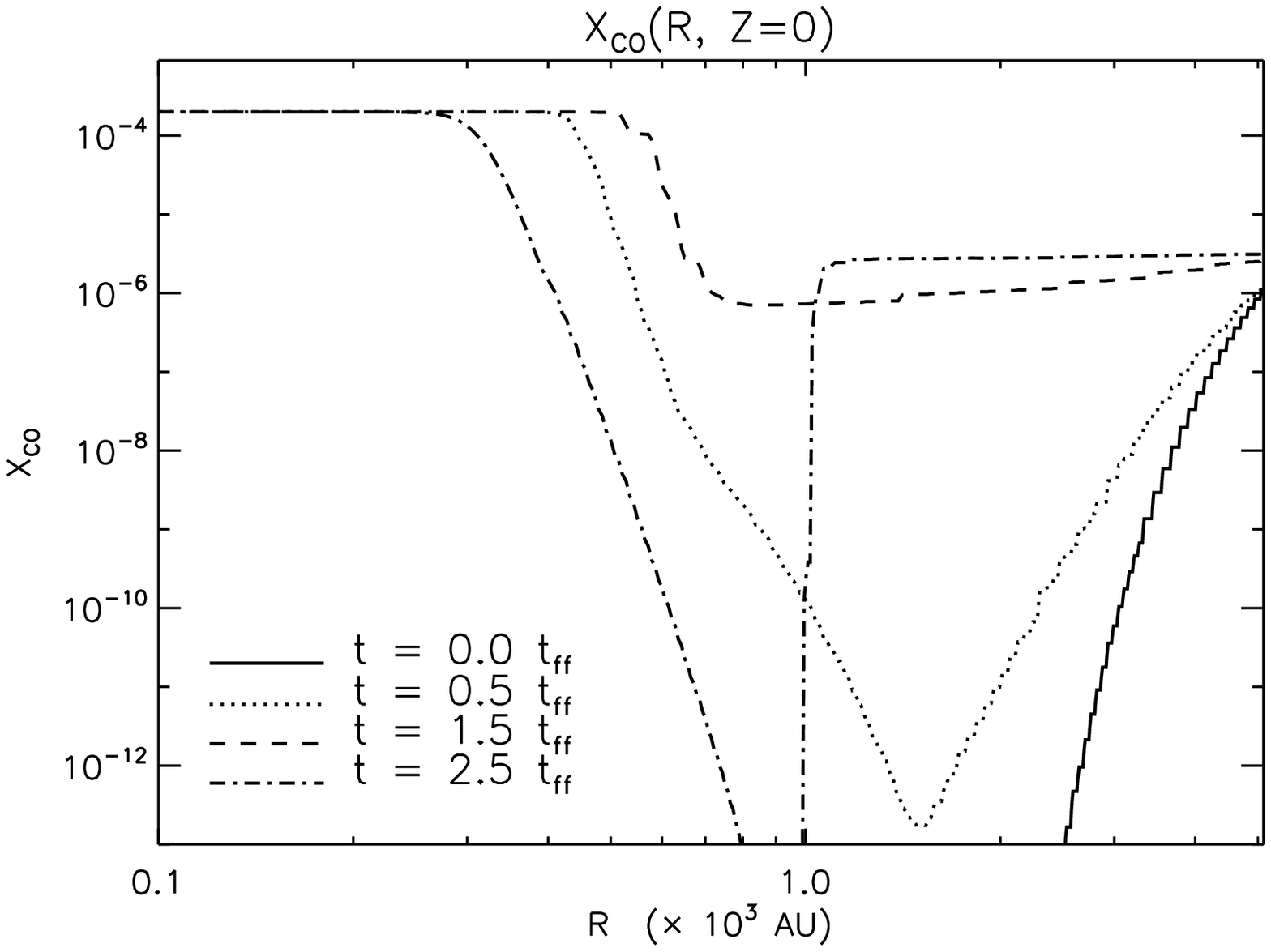}\\
  \includegraphics[width=8.5cm]{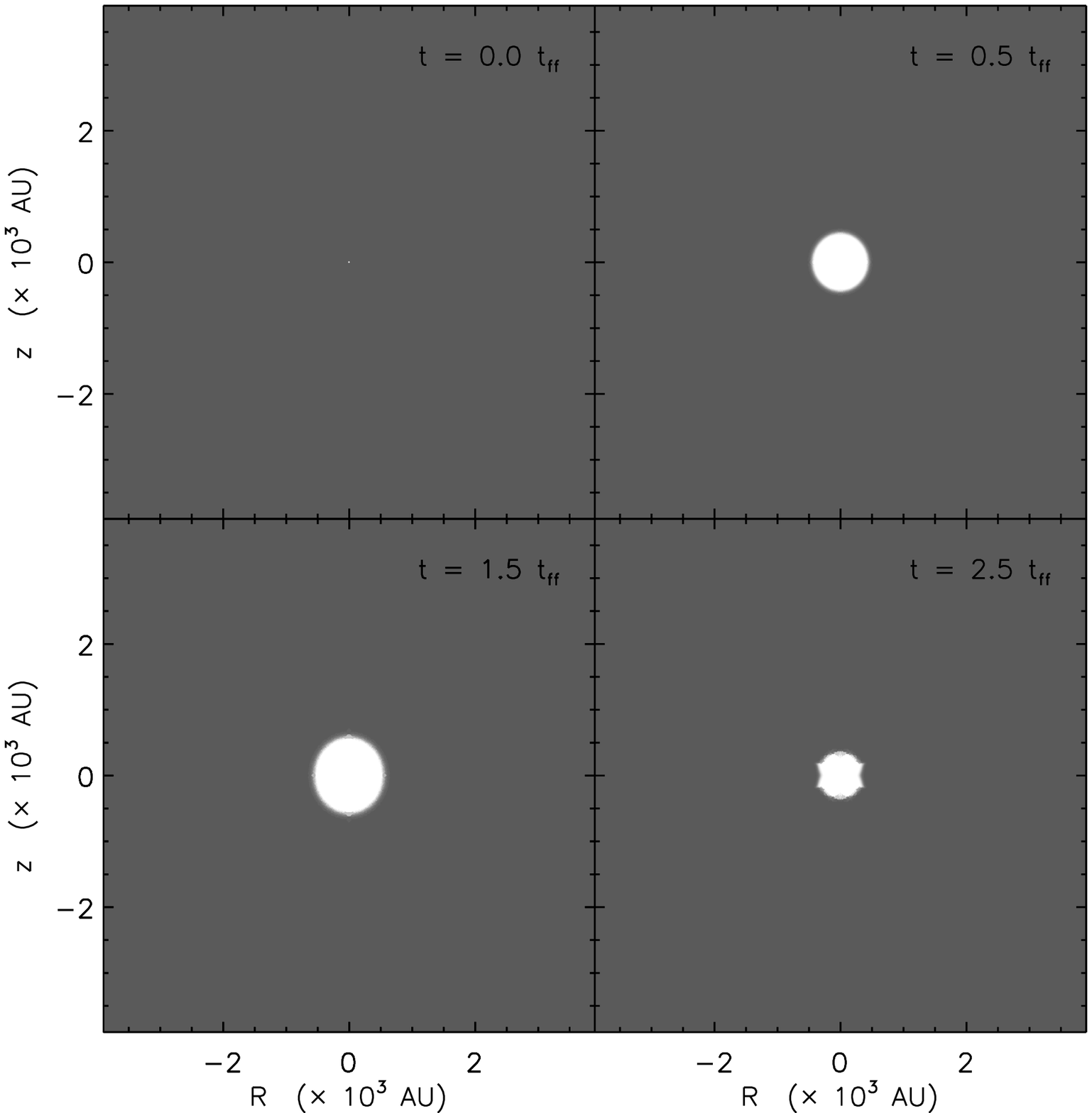}
  \caption{Abundance distribution of Model 3 without cosmic ray desorption
  		   and the default binding energy of 960 K. Otherwise similar to
		   \ref{model1}.}\label{cr}
\end{figure}

Cosmic ray desorption ensures that depletion can never reach 100\% as long as
cosmic rays are able to penetrate the cloud and remove molecules from the ice
matrix. If no other desorption mechanisms were present, dark cold molecular 
clouds would have a very high degree of depletion since thermal desorption is
largely ineffective at low temperatures. However,~\citet{caselli1999} find no
evidence for such high depletion factors and therefore some other desorption 
mechanism (such as for instance cosmic ray desorption) must play a certain role 
in these environments. However, to see exactly how much evaporation cosmic
rays account for, we have calculated Model 3 without the cosmic ray desorption
term. Figure~\ref{cr} show these abundance distributions. The profiles are 
largely similar to Model 3, but the amount of depletion is, as expected, about 
three orders of magnitude higher. 

Including cosmic ray desorption obviously has a big impact on the amount of gas
phase molecules in the cloud, raising the gas-phase abundance significantly. 
However the effect of cosmic ray desorption can be countered by a larger 
binding energy between molecule and grain surface. For example, raising the binding 
energy from 960 K to 1740 K almost entirely compensates the effect of cosmic ray 
desorption, when the depletion is averaged over the entire model.

\begin{figure}
  \includegraphics[width=8.5cm]{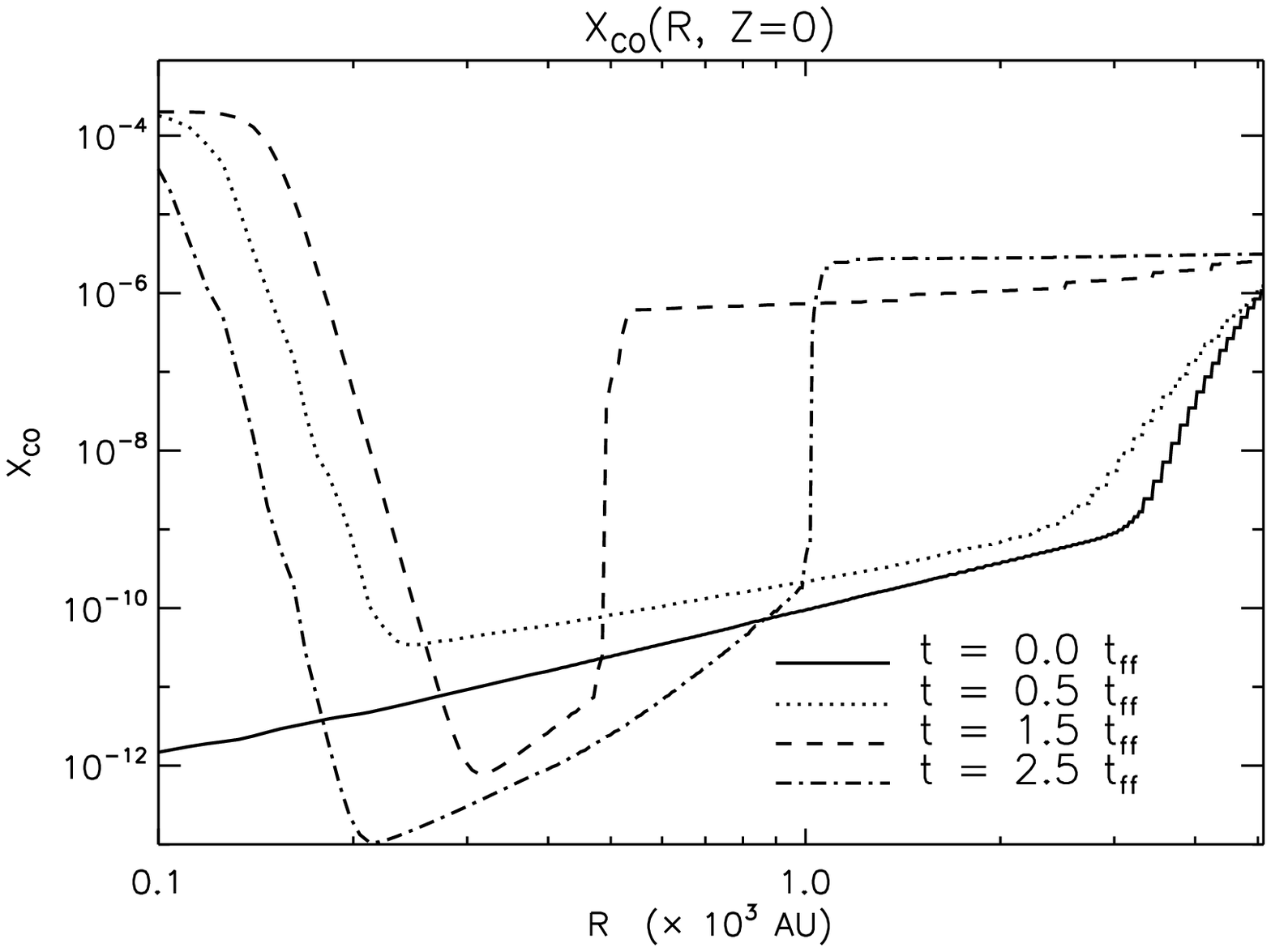}\\
  \includegraphics[width=8.5cm]{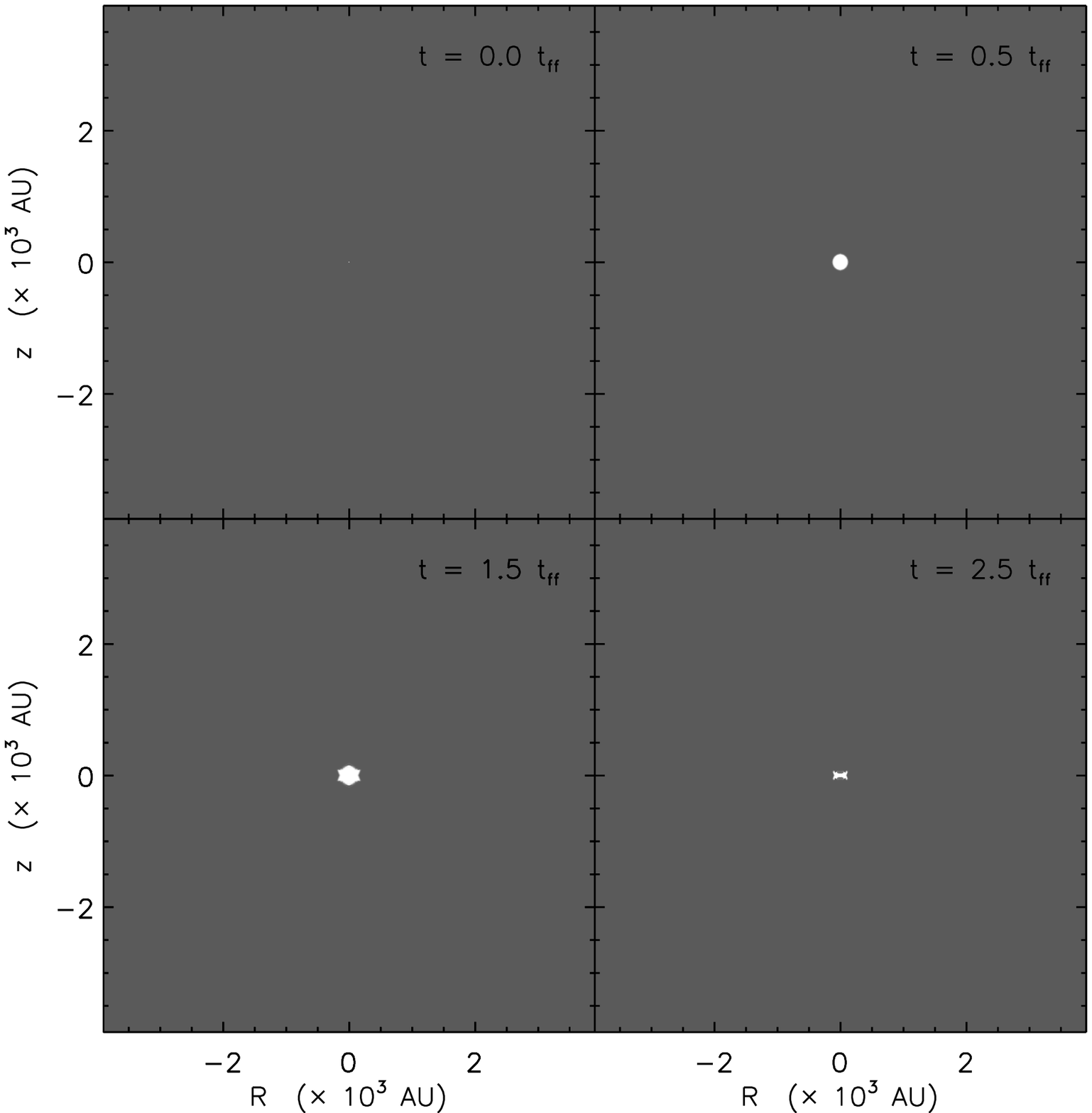}
  \caption{Abundance distribution of Model 3, including cosmic rays, but with a 
  CO binding energy of 1740 K. Otherwise similar to \ref{model1}.}\label{eb}
\end{figure}

An increased binding energy applies when the CO ice mantle is mixed with other 
species, typically water ice or NH$_3$ ice~\citep{fraser2004}. Another way to obtain a spread in 
binding energies is when the ice surfaces are not even, e.g., differently 
layered ices, varying mantle thickness, rough surfaces, etc. If the grains are 
more fractal, with cavities which can trap CO molecules, it will become harder 
to desorb these molecules, effectively raising the binding energy. Laboratory 
experiments of~\citet{collings2004} even showed multiple peaks in the CO 
desorption spectrum, implying a range of binding energies. In Fig.~\ref{eb} it 
can be seen that the inner most evaporation zone is consistently and 
significantly smaller when the binding energy is increased compared to when 
cosmic rays are omitted, suggesting that the two effects can be disentangled by 
high-resolution observations.

\subsection{Comparison to observations}
\begin{figure}
  \includegraphics[width=8.5cm]{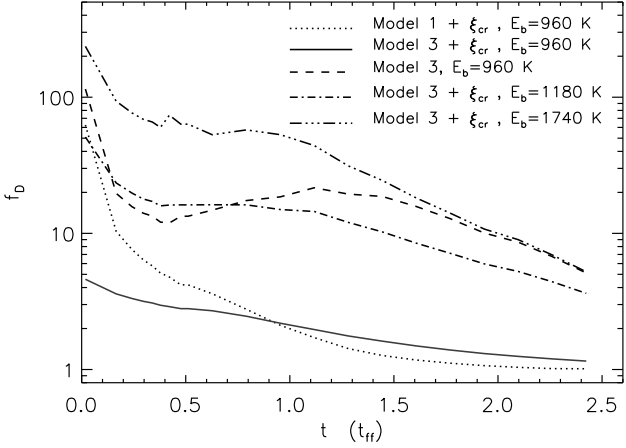}
  \caption{The global fractional depletion in the various models. The small
  		   bump in the curves around 0.5 t$_{ff}$ is due to a change in the 
		   luminosity output of the central source as the disk is formed.
		   }\label{global}
\end{figure}

We compare our results with observations of CO abundances and depletion 
factors. The fractional depletion $f_d$ of CO for the various models are shown 
in Fig.~\ref{global}. The highest depletion factors occur at the end of the 
pre-stellar core phase just before collapse. Observations show that depletion 
factors decrease with decreasing envelope mass~\citep{jorgensen2002}. We indeed 
see the depletion decrease after the core begins to collapse due to CO 
evaporation close to the protostar. At the same time the envelope density 
decreases which also lowers the global fractional depletion, defined as the ratio
of the total number of gas phase CO molecules to H$_2$ molecules in the entire 
model. 

Freeze-out in the disk is most pronounced for Model 3 including cosmic ray
desorption and a high CO binding energy. The same model, but with a binding
energy of 960 K, gives a very low depletion factor at the start of the 
collapse. Although low values of $f_d$ have been measured 
\citep[e.g.,][]{bacmann2002}, most observations point toward a higher value,
$f_d \geq 10$, in pre-stellar cores \citep[e.g.,][]{crapsi2004,bergin2002,
jessop2001}. The low value of $f_d$ we find means that either 
the binding energy is larger than 960 K or that we overestimate the cosmic ray 
desorption rate.  Indeed, \citet{shen2004} showed that the importance of 
cosmic ray heating has been overestimated by previous authors. A third 
possibility is that CO can be converted into other species, resulting in a 
lower abundance, but since we only model CO we cannot quantify this effect.

\begin{figure}
  \includegraphics[width=8.5cm]{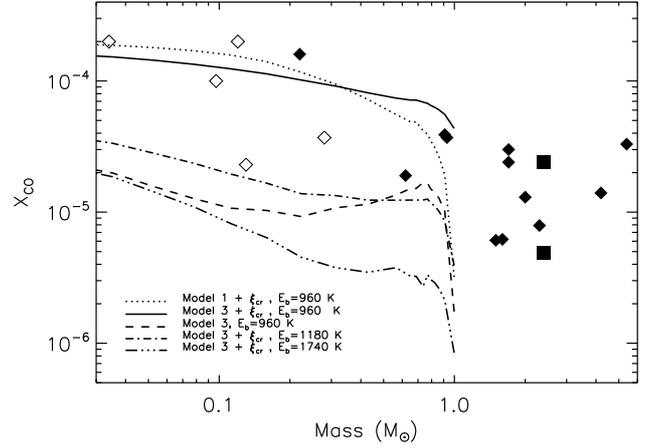}
  \caption{In this plot we show the abundance as function of envelope mass.
           Also shown in this plot are data points taken from~\citet{jorgensen2002,
		   jorgensen2005}. The open symbols are Class I objects, the filled
		   diamonds are Class 0 sources, and the filled squares are pre-stellar 
		   cores.}\label{jes:data}
\end{figure}

As mentioned above, a correlation has been found between the amount of 
depletion and envelope mass. The CO abundance drops when going from a 
pre-stellar core, to a Class 0, to a Class I object. We compare our results to 
the data presented by~\citet{jorgensen2002} who derived global envelope 
abundances for 18 sources. Their result is shown in Fig.~\ref{jes:data}, where 
open symbols are Class I objects, closed diamonds are Class 0 objects, and the 
filled squares are pre-stellar cores. Superposed on the data are curves based 
on our models. To first order our models show the same trend. The observations 
show no evidence for freeze-out in the Class I objects but the number of 
this type of source is small and the derived envelope masses and abundances are 
somewhat uncertain. We also expect some amount of intrinsic scatter due 
to the unique environments and characteristics of each source. Again the drop 
abundance model and Model 3 including cosmic rays and a low binding energy show 
too high CO abundance (or too little depletion). Unfortunately, our simulation 
begins with only one solar mass and therefore we do not catch the more massive 
Class 0 and pre-stellar cores. 

\begin{figure}
  \includegraphics[width=8.5cm]{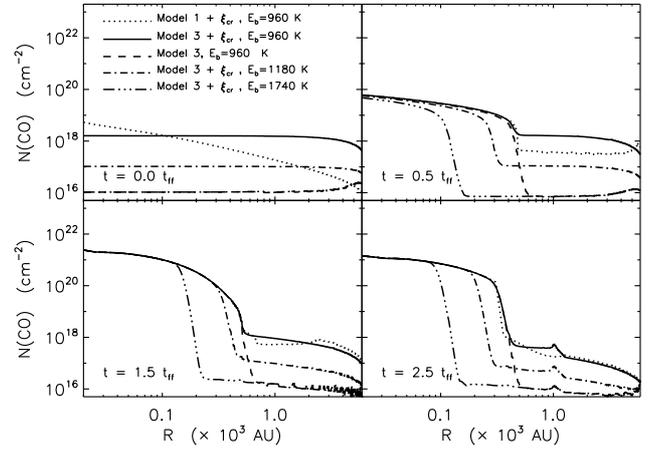}
  \caption{CO gas column densities of the models discussed in this paper.
           These profiles are seen through a pencil beam, and need to be
		   convolved with an appropriate beam in order to be directly
		   compared to observations.}\label{gas:col}
\end{figure}

We can also calculate gas column densities based on our abundance distribution 
models. Vertically integrated column densities, $N_{\rm{CO}} = \int 
n_{\rm{CO}}(z) dz$, are shown at four different free-fall times in 
Fig.~\ref{gas:col}. The models differ mostly on the position of the evaporation 
front as a result of the different binding energies used. The column density 
profiles at t = 0 can be compared to those presented by~\citet{aikawa2001, 
aikawa2003, aikawa2005}. Although our underlying density model is different, 
the column densities are comparable, differing by an order of magnitude or 
less. \citeauthor{aikawa2001} show flat profiles toward the core center or 
slight drops within $\sim$5000 AU, which is the same trend we find with our 
models. The only exception is our drop abundance model (Model 1) which gives an 
increasing column density toward the center. The drop occurs at $\sim$6000 AU, 
and within a 6000 AU radius we have a constant abundance. The $n_{\rm{H}_2}$ 
density decreases quadratically with radius which means that the column density 
of CO increases toward the core center. This is in disagreement with 
observations~\citep[e.g.,][]{Tafalla2004} where flat or decreasing integrated 
intensity profiles are found in core centers, which can be interpreted as a drop 
in the column density.  

Column densities drop around three orders of magnitude from the center toward 
the freeze-out zone in the disk, comparable to~\citet{aikawa1996}, although 
column densities in our results are generally more than one order of magnitude 
higher. This is simply explained by the differences in the adopted disk masses. 
A typical value adopted for disk masses is $10^{-2}$ M$_{\odot}$, while the 
disk in our simulations grows to an unrealistically high mass of about 0.4 
M$_{\odot}$. Also our column density distributions are somewhat more 
complicated as we do not use a simple analytic description of a non-evolving disk. 
\citet{aikawa1996} get drops in column density due to freeze-out in the disk 
around 200 -- 300 AU, similar to our results of 100 -- 500 AU, depending on 
which model we use.

\begin{figure}
  \includegraphics[width=8.5cm]{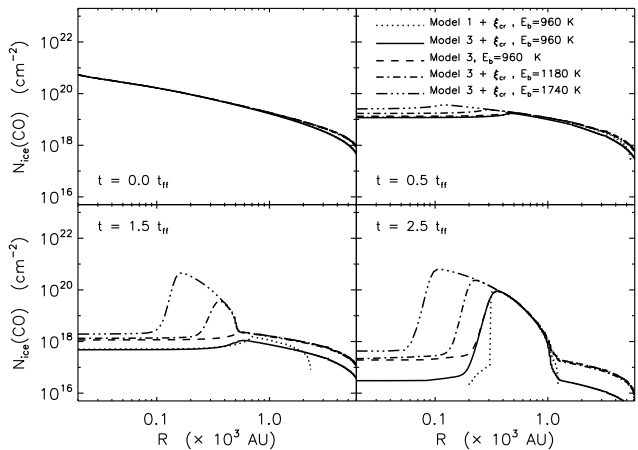}
  \caption{Same as Fig.~\ref{gas:col} but for CO ice column densities. 
  		   }\label{ice:col}
\end{figure}

In a similar way we can calculate the column densities for the CO bound to ice. 
Plots of the ice column densities are shown in Fig.~\ref{ice:col}. Before 
collapse the column density of CO ice increase inwards due to the shorter 
freeze-out timescales there. When the central temperature rises, the column 
density flattens and drops, while in the protoplanetary disk the column density 
sharply increases. Overall, the column density drops in the envelope because of 
the decreasing density (which also results in a lower freeze-out rate). It is 
difficult to compare these results with observations since often cold 
(unrelated) clouds, where CO is frozen out, lie in front of protostars. 
Considering this, observed column densities of CO ice lie often around 
$10^{18}$ cm$^{-2}$ in reasonable agreement with our results. 

\subsection{Emission lines}
Simulated spectral lines are a direct way to test our models since these can 
be compared directly to observed spectra. As mentioned in Section~\ref{model}, 
we use the hydrodynamical time snapshots including the abundance distribution 
mapped onto the original grid as input models for our 2D molecular excitation code 
\emph{RATRAN}. This code solves the level population using an accelerated Monte 
Carlo method, which is then ray-traced to make frequency dependent intensity 
maps. After beam convolution, spectral profiles can be extracted from these and
directly compared to observations. We add a mean field of 0.2 km s$^{-1}$
to the snapshots during the radiation transfer calculations to emulate the 
presence of micro-turbulence.

\begin{figure}
  \includegraphics[width=8.5cm]{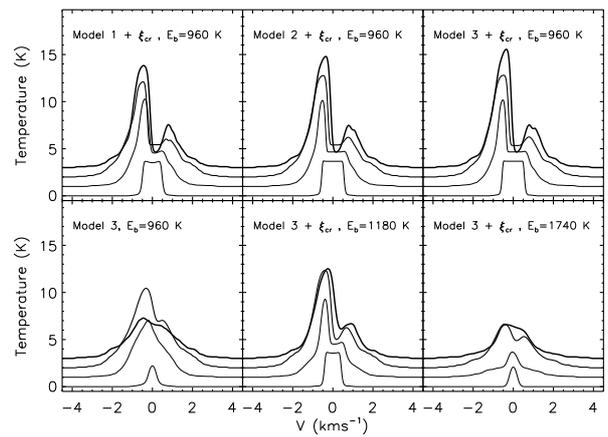}
  \caption{Time series of CO 3--2 spectra. The top three panels show the three
  		   different abundance models. The three lower panels show model 3,
		   including cosmic ray desorption using three different binding 
		   energies. Time is shown by increasing line thickness and the four
		   lines correspond to the four adopted time snapshots (0.0, 0.5, 1.5,
		   and 2.5 t$_{ff}$). Each line has been offset by 0.5 K. }\label{spec1}
\end{figure}

\begin{figure}
  \includegraphics[width=8.5cm]{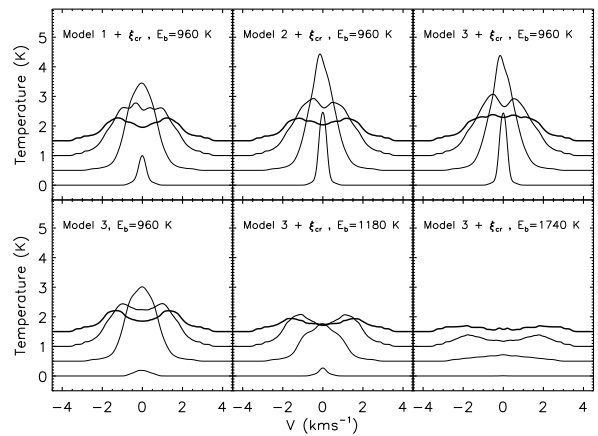}
  \caption{Similar to Fig.~\ref{spec1} but for C$^{18}$O 3--2.}\label{spec2}
\end{figure}

\begin{figure}
  \includegraphics[width=8.5cm]{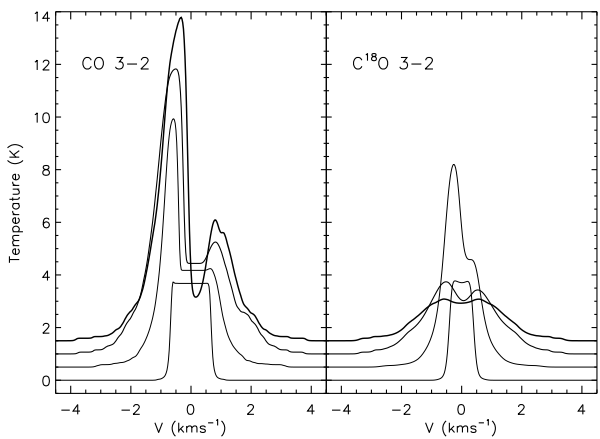}
  \caption{Time series of CO and C$^{18}$O spectra with a constant abundance 
           profile. Time is represented by increasing line thickness.}\label{spec3}
\end{figure}

An in-depth analysis of spectra based on the hydrodynamical simulation is given 
in a separate paper (Brinch et al. \emph{in prep.}) and only example spectra of 
the abundance models derived in this paper are presented here. Figure
\ref{spec1} shows the CO $J=$ 3--2 transition of the four time snap shots which 
have been used in the previous figures. The top three panels correspond to the 
three abundance models including cosmic ray desorption, while the lower three 
panels show spectra based on model 3 without cosmic ray desorption and with 
increasing binding energies. Figure~\ref{spec2} has a similar layout, but for 
the less optically thick isotopologue C$^{18}$O. For comparison, CO and C$^{18}$O 
spectra, calculated with a constant abundance profile are shown in 
Fig.~\ref{spec3}. All spectra are seen under a 90$^\circ$ inclination (edge-on; 
similar to the orientation of the model in Fig.~\ref{hydro-snaps}), and they 
have all been convolved with a 10$''$ beam which is a typical value for 
single-dish sub-millimeter telescopes. A distance of 140 pc (Taurus star 
forming region) is adopted.  We use an isotopic $^{16}$O:$^{18}$O ratio of 560
\citep{wilson1994} to calculate all C$^{18}$O spectra.

First of all, we see that the spectra in the top row of Fig.~\ref{spec1} are 
largely similar to each other and also to the constant abundance profiles in
Fig.~\ref{spec3}. The reason is that CO gets optically thick before the 
abundance goes down and thus they are independent of the underlying abundance 
model. Therefore we cannot use CO spectra to distinguish between the three 
models and we cannot determine the depletion factor using CO. Excluding 
cosmic rays or increasing the binding energy have the effect on the spectra of
lowering the intensity, and in this case CO can be used to constrain the 
models. 

The line profile of the earliest snapshot (the thinnest line) is seen to have a
very boxy shape with a flat line center. This is entirely due to the 
hydrodynamical model, which has not yet evolved from its initial isothermal 
state in the region where the line get optically thick. Basically we probe the 
same temperature through a range of velocities, which gives rise to the flat 
top. The line of the earliest snapshot is also considerably more narrow than
the other lines. This is because the velocities in the model have not yet had
the time to build up.

The C$^{18}$O spectra however are seen to differ considerably from the ones
calculated from a constant abundance model. They also differ, although somewhat
less, from each other when the three abundance models are compared. The reason 
is that the lines are optically thin and therefore a drop in the abundance has
a direct impact on the spectral line, no matter where the drop occurs. 

The spectra shown here are all simulating single-dish measurement, where the
emission of the source has been smeared out in a relatively large beam. If 
instead an interferometer is used, beam sizes better than 1$''$ can be obtained,
which allows us to map the line-of-sight column densities. With such 
observations, it should in principle be possible to map out the abundance 
distributions of optically thin species and constrain the models. 

\section{Discussion}\label{discussion}
The abundance profiles presented here depend on the
adopted hydrodynamical model. While we believe that the overall
behavior of our results are generally applicable, details such as the exact 
timescale and the absolute values of the fractional depletion depend on the
initial conditions of our simulation such as mass and angular momentum and also 
on the particular hydrodynamical scheme which we have used. 

A main concern regarding the hydrodynamics is the internal radiation transfer
module that is used to calculate the temperature structure of the simulation. 
This is done with an approximate method which is known to overestimate the 
dust temperature by a few Kelvins compared to calculation done by a more 
accurate continuum radiation transfer code (Visser, priv.~comm.). While a few 
Kelvin have little or no effect on the hydrodynamics and therefore on the 
evolution of
the cloud, it can certainly make a difference for the evaporation front which
has a strong temperature dependence. While this will affect the detailed 
behavior of the abundances, it will not change the general trends discussed in 
this paper.

Although in this paper we only concerned ourselves with the CO gas-phase 
abundance, as mentioned in Sect.~\ref{model}, our method can be easily extended 
to include other species and even surface reactions. However, running the 
full chemical network is considerably more computationally demanding. Whereas
the CO model runs in a few hours on a standard desktop computer using $9 \times
10^5$ trace particles, a test run of the full chemical network showed that it
takes a factor of 100 longer to run using a factor of 1000 less trace 
particles. Nevertheless, such large scale chemical models present a powerful
way to obtain realistic abundance profiles for hydrodynamical simulations which
instruments like the Atacama Large Millimeter Array (ALMA) can test at high
spatial resolution.

\section{Conclusion}\label{summary}
In this paper we have investigated different scenarios for the depletion of CO
in a collapsing cloud. We have  used a hydrodynamical code to describe the
dynamical evolution of the cloud and the chemistry has been solved using a 
trace particle approach.

Of the various models shown in Sect.~\ref{results} the one which exclude 
cosmic ray desorption and the ones using higher binding energies differ the 
most. The drop abundance model provides a good approximation to Model 2 and 3. 
Therefore, if it is possible to fit a drop profile to observations, as was done 
by~\citet{jorgensen2005}, it should be possible to run a hydrodynamical
simulation with customized initial conditions, so that it reproduces this 
profile and thereby putting an observational constraint on the hydrodynamics. 
Also in cases where an analytical model is used, using the drop abundance 
approximation is quite good.

We furthermore conclude that the freeze-out chemistry has little effect on
optically thick spectral lines, especially when observed in low resolution. It
is therefore not entirely unreasonable to model such lines using a constant
abundance model disregarding the chemistry. However, one should be careful
when using optically thin lines to determine which average gas phase abundance to use
for a constant abundance model, since the average abundances derived from such
lines are indeed depending on the freeze-out chemistry. The maximum
discrepancy between the flat abundance profile model (Fig.~\ref{spec3}) and
the freeze-out models shown in Fig.~\ref{spec2} is a factor of three in the
integrated intensity. Such a change in the average abundance does not affect the 
optically thick lines a lot, so depending on the accuracy needed, using a flat 
abundance distribution to model data in which freeze-out is present may not be 
entirely unreasonable.

Finally, the effect of the chemical depletion is considerably enhanced when
the resolution of the observations is increased, making it more important to
get the abundance profiles right. This will especially be important for future
ALMA observations (with a resolution of $\sim$ 0.01$''$) where the level of
detail will be so great that getting the exact abundances will be crucial for 
interpreting the data.\\

\noindent \emph{Acknowledgments} The authors would like to thank the anonymous 
referee for thorough and useful comments. CB is partially supported by the 
European Commission through the FP6 - Marie Curie Early Stage Researcher Training 
programme. The research of MRH is supported through a VIDI grant from the 
Netherlands Organization for Scientific Research.\\

\bibliographystyle{aa}
\bibliography{/home/brinch/papers/references}
\end{document}